\documentclass[12pt,english]{article}
\usepackage[T1]{fontenc}
\usepackage[latin1]{inputenc}
\usepackage{babel}

\makeatletter

\newcommand{\LyX}{L\kern-.1667em\lower.25em\hbox{Y}\kern-.125emX\spacefactor1000}
\let\SF@@footnote\footnote
\def\footnote{\ifx\protect\@typeset@protect
    \expandafter\SF@@footnote
  \else
    \expandafter\SF@gobble@opt
  \fi
}
\expandafter\def\csname SF@gobble@opt \endcsname{\@ifnextchar[
  \SF@gobble@twobracket
  \@gobble
}
\edef\SF@gobble@opt{\noexpand\protect
  \expandafter\noexpand\csname SF@gobble@opt \endcsname}
\def\SF@gobble@twobracket[#1]#2{}

\newcommand{\lyxaddress}[1]{
  \par {\raggedright #1 
  \vspace{1.4em}
  \noindent\par}
}

\usepackage{espcrc1}

\makeatother

\begin{document}

\title{Solution of the Faddeev-Yakubovsky equations using realistic NN and 3N interactions}

\author{A. Nogga\protect\( ^{\rm a}\protect \), H.Kamada\protect\( ^{\rm a}\protect \),
and W.Glöckle\protect\( ^{\rm a}\protect \)}

\maketitle

\lyxaddress{\protect\( ^{\rm a}\protect \)Institut für theoretische Physik II, Ruhr-Universität,
D-44780 Bochum, Germany}

\begin{abstract}
We solve the Faddeev-Yakubovsky equations for 3N and 4N bound states based on
the most modern realistic nucleon-nucleon interactions. We include different
realistic 3N forces. It is shown that all 3N force models can remove the underbinding
of the triton and \( \alpha  \)-particle which one obtains with existing NN
interactions. The agreement of theoretical predictions and the experimental
binding energy is quite good and there is little room left for the action of
four-nucleon forces in the \( \alpha  \)-particle. The effect of 3N forces
on the wave function is investigated.
\end{abstract}

\section{Introduction}

In recent years it has become possible to solve rigorously the 4N Faddeev-Yakubovsky
equations in momentum space \cite{kamada92,nogga00}. Using parallel computers
one can achieve converged solutions for realistic nuclear Hamiltonians. There
are several realistic interactions available which can describe the nuclear
two-body problem with a very high accuracy. These are the potentials of the
Nijmegen group Nijm I, II and 93 \cite{nijm93}, the CD-Bonn \cite{cdbonn}
and the AV18 interaction \cite{av18}. All these interactions account for charge
independence breaking (CIB). The CD-Bonn and AV18 additionally distinguish pp
and nn forces. It is well known that nonrelativistic calculations based on these
potentials lead to an underbinding of 500-800 keV in the 3N system. Two additional
dynamical ingredients should cure this underbinding: relativistic corrections
and three-nucleon forces (3NF). In this contribution we would like to address
the question whether 3NF's can solve the underbinding problem in both the 3N
and 4N system and investigate the dependence of binding energies and wave functions
on the NN and 3N force model.

\section{Binding energies}

In Table ~\ref{Tab:1} we compare our theoretical predictions for various nuclear
potentials to the experimental binding energies. The results depend on the NN
force model, but the experimental binding energy of the triton is always underpredicted
by 500-800 keV for all nuclear potentials. The weakest binding is predicted
by the three local potentials AV18, Nijm 93 and Nijm II, whereas the strongly
non local CD-Bonn gives \( \approx 300 \) keV more binding. The non local potentials
are less repulsive for small distances and their short range correlations are
softer. This is visible in their smaller kinetic energy predictions \cite{nogga97}.
A linear correlation found in \cite{tjon75} for simple forces holds also in
case of the modern realistic forces including charge independence and symmetry
breaking \cite{nogga00}. 

\begin{table}

\caption{{\footnotesize \label{Tab:1}3N and 4N binding energies for various NN potentials
with expectation values \protect\( T\protect \) of the kinetic energy. }\footnotesize }

{\centering \begin{tabular}{l|cc|cc}
{\footnotesize }&
\multicolumn{2}{|c|}{ {\footnotesize \( ^{3} \)H}}&
\multicolumn{2}{|c}{{\footnotesize \( ^{4} \)He }}\\
\hline 
{\footnotesize Potential}&
{\footnotesize \( E_{B} \){[}MeV{]}}&
{\footnotesize \( T \){[}MeV{]}}&
{\footnotesize \( E_{B} \){[}MeV{]}}&
{\footnotesize \( T \){[}MeV{]}}\\
\hline 
{\footnotesize CD-Bonn}&
{\footnotesize -8.012}&
{\footnotesize 37.42}&
{\footnotesize -26.26}&
{\footnotesize 77.15}\\
{\footnotesize AV18}&
{\footnotesize -7.623}&
{\footnotesize 46.73}&
{\footnotesize -24.28}&
{\footnotesize 97.83}\\
{\footnotesize Nijm I}&
{\footnotesize -7.736}&
{\footnotesize 40.73}&
{\footnotesize -24.98}&
{\footnotesize 84.19}\\
{\footnotesize Nijm II}&
{\footnotesize -7.654}&
{\footnotesize 47.51}&
{\footnotesize -24.56}&
{\footnotesize 100.31}\\
{\footnotesize Nijm 93}&
{\footnotesize -7.661}&
{\footnotesize 45.60}&
{\footnotesize -24.53}&
{\footnotesize 95.34}\\
\hline 
{\footnotesize exp.}&
{\footnotesize -8.48}&
{\footnotesize -}&
{\footnotesize -28.30}&
{\footnotesize -}\\
\end{tabular}\footnotesize \par}
\end{table} 

The inclusion of 3NF's cannot result in parameter free predictions for the 3N
and 4N system because all realistic potentials are not accompanied by theoretically
founded 3NF's based on the same fundamental dynamical assumptions. Therefore
we use the Tucson-Melbourne(TM) 2\( \pi  \) 3NF \cite{coon79,coon81} and adjust
the cutoff parameter in the NN\( \pi  \) form factors individually for each
NN interaction insuring that the resulting 3N Hamiltonians reproduce the experimental
triton binding energy. The Nijmegen forces do not specify a neutron-neutron
force. In this case we used \( ^{3} \)He for the adjustment. The resulting
cutoff values are given in Table~\ref{Tab:2} . In \cite{friar99} a modified
version of the TM force was suggested which at least does not violate chiral
symmetry. We call it TM' and adjusted it to the AV18 force in the same manner.
Additionally, we included the Urbana-IX 3N interaction \cite{pudliner97}. 

As shown in Table \ref{Tab:2} all force combinations overpredict the \( \alpha  \)-particle
binding energy. However the overbinding is rather small. This indicates that
the contribution of 4N forces is small in the \( \alpha  \)-particle. Of course,
on this phenomenological level the addition of more structures to the 3NF (for
example the \( \pi -\rho  \) exchange) might change this situation.

\begin{table}

\caption{\label{Tab:2}{\footnotesize Cutoff parameters \protect\( \Lambda \protect \),
3N and 4N binding energies for various force combinations with expectation values
\protect\( T\protect \) of the kinetic energy. }\footnotesize }

{\centering \begin{tabular}{l|c|cc|cc}
{\footnotesize }&
 {\footnotesize }&
\multicolumn{2}{|c|}{{\footnotesize \( ^{3} \)H}}&
\multicolumn{2}{|c}{{\footnotesize \( ^{4} \)He}}\\
\hline 
{\footnotesize Potential}&
{\footnotesize \( \Lambda  \){[}\( m_{\pi } \){]}}&
{\footnotesize \( E_{B} \){[}MeV{]}}&
{\footnotesize \( T \){[}MeV{]}}&
{\footnotesize \( E_{B} \){[}MeV{]}}&
{\footnotesize \( T \){[}MeV{]}}\\
\hline 
{\footnotesize CD-Bonn+TM}&
{\footnotesize 4.784}&
{\footnotesize -8.480}&
{\footnotesize 39.10}&
{\footnotesize -29.15}&
{\footnotesize 83.92}\\
{\footnotesize AV18+TM}&
{\footnotesize 5.156}&
{\footnotesize -8.476}&
{\footnotesize 50.76}&
{\footnotesize -28.84}&
{\footnotesize 111.84}\\
{\footnotesize AV18+TM'}&
{\footnotesize 4.756}&
{\footnotesize -8.444}&
{\footnotesize 50.55}&
{\footnotesize -28.36}&
{\footnotesize 110.14}\\
{\footnotesize AV18+Urb-IX}&
{\footnotesize -}&
{\footnotesize -8.478}&
{\footnotesize 51.28}&
{\footnotesize -28.50}&
{\footnotesize 113.21}\\
{\footnotesize Nijm I+TM}&
{\footnotesize 5.035}&
{\footnotesize -8.392}&
{\footnotesize 43.35}&
{\footnotesize -28.60}&
{\footnotesize 93.58}\\
{\footnotesize Nijm II+TM}&
{\footnotesize 4.975}&
{\footnotesize -8.386}&
{\footnotesize 51.02}&
{\footnotesize -28.54}&
{\footnotesize 113.09}\\
\hline 
{\footnotesize exp.}&
{\footnotesize -}&
{\footnotesize -8.48}&
{\footnotesize -}&
{\footnotesize -28.30}&
{\footnotesize -}\\
\end{tabular}\footnotesize \par}
\end{table}

\section{Wave function properties}

In this section we address the question whether 3NF's have got any effects on
the wave functions. Because the effects in the 3N and 4N systems are very similar
and are more pronounced in the 4N system, we restrict ourselves to the \( \alpha  \)-particle. 

It is well known that 3NF's are not treatable by perturbation theory \cite{bomelburg86,friar88}.
Therefore the inclusion of 3NF's change the wave function. Surprisingly for
us we found no significant differences in the nuclear momentum distributions
and NN correlations comparing wave functions generated with and without 3NF.
This holds for all force combinations we used. 

To reveal the differences in the wave functions, we calculated expectation values
of the 3NF operators. In Table \ref{Tab:3} we compare the expectation values
of the different 3NF's obtained with wave functions which we generated using
the AV18, AV18+TM, AV18+TM' and AV18+Urbana IX interactions. The results of
the wave function generated from the AV18 NN interaction alone are generally
much smaller in magnitude than the values obtained from the ``full'' wave
functions. Clearly the expectation values differ from the gain in binding energy
one obtains with 3NF's (see Table \ref{Tab:1} and \ref{Tab:2}). Therefore
the 3NF operators are sensitive to the differences in the wave functions. \begin{table}

\caption{\label{Tab:3}{\footnotesize Expectation values of the TM, TM' and Urbana-IX
force calculated with wave functions we generated using the AV18, AV18+TM,AV18+TM'
and AV18+Urbana\protect\( \, \protect \)IX force combinations. The cutoff values
of TM and TM' are adjusted to the AV18 potential and given in Table \ref{Tab:2}.}\footnotesize }

\vspace{0.3cm}
{\centering \begin{tabular}{l|c|c|c}
{\footnotesize }&
{\footnotesize <TM>}&
{\footnotesize <TM'>}&
{\footnotesize <Urb-IX>}\\
\hline 
{\footnotesize \( \Psi  \)(AV18)}&
{\footnotesize -1.60}&
{\footnotesize -3.13}&
{\footnotesize -2.64}\\
{\footnotesize \( \Psi  \)(AV18+TM)}&
\textbf{\footnotesize -8.03}{\footnotesize }&
{\footnotesize -4.76}&
{\footnotesize -4.20}\\
{\footnotesize \( \Psi  \)(AV18+TM')}&
{\footnotesize -2.99}&
\textbf{\footnotesize -5.17}{\footnotesize }&
{\footnotesize -4.98}\\
{\footnotesize \( \Psi  \)(AV18+Urb-IX)}&
{\footnotesize -3.31}&
{\footnotesize -5.48}&
\textbf{\footnotesize -5.94}\\
\end{tabular}\footnotesize \par}
\vspace{0.3cm}
\end{table} 

In the next step we restrict the wave functions to their S,P and D-wave components
and calculate the transition matrix elements of the Urbana IX and TM 3NF. The
most important matrix elements are summarized in Table \ref{Tab:4}. The contribution
of the different partial waves is very 3NF dependent. Whereas the repulsive
part of the Urbana interaction is visible in the diagonal matrix elements, the
TM force is strongly attractive in the S-S matrix element. The usage of the
wave function generated with TM increases the attraction in the S-S and S-P
transition. Most of the attraction in the Urbana force is generated by the S-D
transition. In contrast these particular partial waves are extremely wave function
dependent for the TM force. The results indicate that though the binding energies
are similar both 3NF operators act very differently on the wave functions of
the \( \alpha  \)-particle. \begin{table}

\caption{\label{Tab:4}{\footnotesize Contribution of different parts of the wave function
to the Urbana-IX and TM expectation values. In the first row the 3N interaction
is chosen. In the second row the Hamiltonian, which is used to generate the
wave function, is fixed and in the first column the part of the wave function
on the right and left is given. Transition matrix elements are multiplied by
2. The cutoff of TM is adjusted to the AV18 potential and given in Table \ref{Tab:2}.}\footnotesize }

\vspace{0.3cm}
{\centering \begin{tabular}{l|rr|rr}
{\footnotesize }&
\multicolumn{2}{|c|}{{\footnotesize Urbana IX}}&
\multicolumn{2}{|c}{{\footnotesize TM }}\\
\hline 
{\footnotesize }&
{\footnotesize \( \Psi  \)(AV18)}&
{\footnotesize \( \Psi  \)(AV18+UrbIX)}&
{\footnotesize \( \Psi  \)(AV18)}&
{\footnotesize \( \Psi  \)(AV18+TM)}\\
\hline 
{\footnotesize S-S}&
{\footnotesize +3.16}&
{\footnotesize +2.74}&
{\footnotesize -2.34}&
{\footnotesize -4.09}\\
{\footnotesize S-P}&
{\footnotesize -0.96}&
{\footnotesize -2.10}&
{\footnotesize -1.22}&
{\footnotesize -3.56}\\
{\footnotesize S-D}&
{\footnotesize -5.44}&
{\footnotesize -7.46}&
{\footnotesize +2.08}&
{\footnotesize -0.14}\\
{\footnotesize D-D}&
{\footnotesize +0.59}&
{\footnotesize +0.85}&
{\footnotesize +0.01}&
{\footnotesize +0.06}\\
\end{tabular}\footnotesize \par}
\vspace{0.3cm}
\end{table}

\section{Conclusions}

We presented 3N and 4N binding energies using the most modern nuclear interactions.
We showed that the underbinding of few-nucleon bound states, which is predicted
by purely two-nucleon interactions, is solved adding a 3NF. This holds for all
combinations of 3NF and NN interaction and indicates that 4N forces, at least
in the \( \alpha  \)-particle, are small. A look to NN correlations in the
wave functions generated from different nuclear Hamiltonians reveals that all
investigated 3NF's do not change the short range correlations in the 3N and
4N system. Because we found at the same time that the expectation values of
different 3N interactions are strongly dependent on the used wave function,
we conclude that these wave functions have got very different 3N correlations.
The next task of our study is to pin down these differences and visualize them.
This might help to clarify the effects of 3NF's in the 3N and 4N system. 

This work was supported financially by the Deutsche Forschungsgemeinschaft (A.N.
and H.K.). The numerical calculations were performed on the Cray T3E of the
NIC in Jülich.

\bibliographystyle{prsty}
\bibliography{literatur}

\end{document}